# Si microring resonator crossbar array for on-chip inference and training of optical neural network

Shuhei Ohno[1*], Kasidit Toprasertpong[1], Shinichi Takagi[1], and Mitsuru Takenaka[1**]

[1]The University of Tokyo, 7-3-1 Hongo, Bunkyo-ku, Tokyo 113-8656, Japan

*E-mail: ohno@mosfet.t.u-tokyo.ac.jp

**E-mail: takenaka@mosfet.t.u-tokyo.ac.jp

**Deep learning is one of the most advancing technologies in various fields. Facing the limits of the current electronics platform, optical neural networks (ONNs) based on Si programmable photonic integrated circuits (PICs) have attracted considerable attention as a novel deep learning scheme with optical-domain matrix-vector multiplication (MVM). However, most of the proposed Si programmable PICs for ONNs have several drawbacks such as low scalability, high power consumption, and lack of frameworks for training. To address these issues, we have proposed a microring resonator (MRR) crossbar array as a Si programmable PIC for an ONN. In this article, we present a prototype of a fully integrated 4 × 4 MRR crossbar array and demonstrated a simple MVM and classification task. Moreover, we propose on-chip backpropagation using the transpose matrix operation of the MRR crossbar array, enabling the on-chip training of the ONN. The proposed ONN scheme can establish a scalable, power-efficient deep learning accelerator for applications in both inference and training tasks.**

Introduction

In the era of big data, artificial intelligence (AI) based on machine learning plays an important role in obtaining meaningful information from numerous and complex data. In particular, deep learning is one of the most powerful tools in machine learning[1]. Deep learning has several forms represented by convolutional neural networks (CNNs) and recurrent neural networks (RNNs), and it has a wide range of applications such as image recognition, translation, medical diagnostics, and so on. A neural network consists of multiple matrix-vector multiplications (MVMs), which cost long computation time and high energy consumption. Thus, it goes without saying that deep learning technologies cannot be improved without the development of specific hardware architectures based on a complementary metal-oxide-semiconductor (CMOS) platform[2,3]. Simultaneously with improving the electronics hardware for deep learning, novel deep learning schemes called optical neural networks (ONNs) have been proposed for several methods[4,5,6]. The computation speed of MVM can be improved by using optical signals for computation, without taking into account the physical limitation of current CMOS platform[7]. In particular, ONNs based on a Si photonics platform have attracted a great deal of attention. Si photonics allows us to fabricate Si programmable photonic integrated circuits (PICs) consisting of scalable, low-loss Si waveguides and optical phase shifters, which can be used for optical-domain MVM with arbitrary matrices. There are mainly two types of Si programmable PIC for ONNs. One type consists of Mach–Zehnder interferometer (MZI) switches lined up in a mesh (meshed-MZI)[8,9,10]. The other type has cascaded add-drop microring resonators (MRRs)[11] with wavelength division multiplexing (WDM) optical signal, which is called the weight bank[12]. In particular, prototype processors for ONNs based on meshed-MZI have been developed in several start-ups[13,14]. However, several issues remain in these Si programmable PICs. In the case of meshed-MZI, the large size of a MZI hinders the scalability of PICs. In addition, complex computation is required to set a matrix to a PIC owing to the meshed structure[15,16]. On the other hand, the weight bank has inner loops inside a PIC, making it difficult to expand its circuit size[12]. Moreover, for both types of PIC, we cannot perform the backpropagation algorithm in the PICs, which accelerates the training of a neural network[17]. Thus, ex situ training or brute force gradient computation in a PIC is required when using an ONN based on conventional Si programmable PICs[8], which wastes the computation speed of the PIC. In previous works, the backpropagation algorithm based on an adjoint method has been proposed and numerically demonstrated[18]; however, it is difficult to perform on-chip demonstration because the optical intensity of each phase shifter in the PIC needs to be measured. An ONN using a photorefractive crystal has also been proposed for on-chip backpropagation[19]; however, the method of programming a photorefractive crystal is still not clear. In this paper, to address these issues, we propose and demonstrate both the on-chip inference and training of an ONN based on a MRR crossbar array. Figure 1(a) shows a schematic of a MRR crossbar array for 4 × 4 MVM. Regarding the MVM procedure for an inference task, we inject WDM optical signals corresponding to input vector $\boldsymbol{x}$ from Ports 1–4, which can be expressed as Forward signals. The different colors of the MRRs in Fig. 1(a) correspond to the different wavelengths of the WDM optical signals. An add-drop MRR with a phase shifter works as a multiplier for each wavelength. By tuning each phase shifter, we can adjust each element of a weight matrix $\boldsymbol{W}$ for the multiply operation. Since the dropped output of the MRR can be directly monitored at each wavelength of the WDM signal, it is more straightforward to set the elements of $\boldsymbol{W}$ than when using meshed-MZI. Finally, the add operation is carried out by converting the dropped WDM signal from multiple MRRs into an electrical signal with a photodetector (PD) simultaneously. Furthermore, by injecting optical signals from Ports 5–8, which can be defined as Backward signals, we can perform MVM with the transpose matrix for the matrix calculated with Forward signals without tuning the matrix elements. This transpose matrix operation is applicable to on-chip backpropagation using the auxiliary error vector $\boldsymbol{\delta}$. Thus, the MRR crossbar array has potential use for both on-chip inference and training tasks for ONNs. In previous works, we demonstrated the principle of MVM using an MRR crossbar array consisting of up to nine MRRs[20,21]. In this work, we demonstrate a fully integrated MRR crossbar array with optical modulator arrays for Forward and Backward signals and Ge PD arrays. In addition to the inference, we successfully show the feasibility of the on-chip backpropagation using a MRR crossbar array, which can be used for on-chip training.

**Results**

**Chip layout.** Figure 1(b) shows a plan-view image of the 4×4 MRR crossbar array fabricated on a Si photonics platform. Its total chip size is 4.3 mm × 2.0 mm. A single-mode Si strip waveguide with a height of 210 nm and a width of 400 nm was used to design this Si programmable PIC operating at a wavelength of 1550 nm. Sixteen MRRs are integrated in a crossbar array configuration. The gap and radius of each MRR are 200 nm and 10 μm, respectively. Thermo-optic (TO) phase shifters with TiN heaters are implemented to tune the drop signal of the MRRs. A multimode structure including four ellipses is employed at each intersection of the MRR crossbar array to eliminate its insertion loss and crosstalk[22]. Ge PD[23,24] arrays are integrated at the outputs of the MRR crossbar array to detect drop signals. To generate Forward and Backward signals, MZI modulator arrays are implemented on the left and upper right sides of the MRR crossbar array. TO phase shifters are also used for the MZI modulators. WDM optical signals are injected to the input waveguides as Forward and Backward signals through a spot-size converter and equivalently distributed into the MZI modulator arrays by cascaded 3 dB couplers based on a multimode interference (MMI) coupler. The measurements are described in Method and Supplementary Section I, II.

**Output of MRR crossbar array.** First, we injected a single wavelength optical signal from a single port to evaluate the wavelength characteristics of the MRR crossbar array by sweeping the wavelength of the signal. Before the measurement, we roughly tuned the drop peaks of the MRRs with injection currents of the phase shifters to fit the assigned wavelengths of the WDM optical signals. Figure 1(c) shows the obtained output spectra of drop signals from the MRR crossbar array when an optical signal is injected from Port 1. Different colors correspond to the photocurrent obtained from different PDs in Fig. 1(a). The four lines in the figure correspond to wavelengths assigned to the WDM optical signals, $\lambda_1$, $\lambda_2$, $\lambda_3$, and $\lambda_4$. Since all resonance peaks are separated from each other, the matrix elements can be independently programmed to perform MVM. Small noises are found in the characteristics of the MRR owing to the imperfect fabrication of the MZI modulators, which can be eliminated by optimizing their design. Details of the MRR characteristics are also discussed in Supplementary Section III. Next, we adjusted the drop signal of the MRR crossbar array to obtain the desired matrix for MVM with a Forward signal. Due to the thermal crosstalk between the MRRs, it is difficult to control the MRR crossbar array by directly tuning the injection current of the phase shifters. To resolve this issue, we applied feedback control to the MRR crossbar array as follows[25,26]. A WDM optical signal was injected to an input port as the Forward signal as shown in Fig. 2(a). First, the intensities of all elements of the Forward signal vector were set to 1. Then, the drop outputs of the MRRs were adjusted to the target intensities by tuning the phase shifters with the feedback control. This control method has been demonstrated in several related works and is robust to PIC noise. Figures 2(b)–(d) show output currents of PDs 1–4 after the optimization of the phase shifter when Forward signal set to 1 was input to Ports 1–4 of the MRR crossbar array, exhibiting three representative 4 × 4 matrices where the 1 and 0 states in the matrix elements were expressed by 9 and 5 μA of the Ge PDs, respectively. Owing to the feedback calibration, arbitrary matrices were successfully expressed, showing the feasibility of the multiplication of a matrix $\boldsymbol{W}$ and an input vector $\boldsymbol{x}$ for the inference task using the MRR crossbar array. Furthermore, we examined on-chip backpropagation in the MRR crossbar array, which enables the multiplication between a transpose matrix $\boldsymbol{W^T}$ and an auxiliary error vector $\boldsymbol{\delta}$ for learning tasks. After the feedback control for Forward signals shown in Figs. 2(b)–(d), the backward signal $\boldsymbol{\delta}$ was injected into the MRR crossbar array from the backward direction and the output currents of PDs 5–8 were measured to evaluate the transpose matrices. Figure 2(e) shows the route of an optical signal in transpose matrix operation. The output of the PIC becomes smaller as the route in the MRR crossbar array becomes longer due to the attenuation of its waveguide and crossing intersection. Thus, we corrected the output with eight parameters corresponding to four input ports and four PDs, which can be implemented in this system. To find these parameters, non-negative matrix factorization based on an alternating least squares algorithm was carried out. The correction is discussed in detail in Supplementary Section IV. Figures 2(f)–(h) show the results of the transpose matrix operation using the MRR crossbar array, which correspond to the transposes of the matrices in Fig. 2(b)–(d), respectively. The transpose outputs of these matrices were successfully obtained, although there was still a large noise for calculation, which was attributable to the small wavelength-dependent noise of the MZI switches shown in Fig. 1(c).

**Demonstration of inference task.** To evaluate the feasibility of the inference task, we implemented an ONN shown in Fig. 3(a), which includes one hidden layer, with the MRR crossbar array and demonstrated the inference of three species of iris flower using the Iris dataset. Figure 3(b) shows a schematic of the demonstration flow. In the ONN, the ReLU function was employed as the activation function. As shown in the figure, MVM using the MRR crossbar array was carried out twice at a single layer of the ONN for calculation including negative values. By using the offset matrix, all elements of which are the same, we can suppress the circuit size of the MRR crossbar array for MVM including negative values. This method is also discussed in detail in Supplementary Section V. For MVM using an $N \times N$ matrix, the size of the matrix for the offset calculation is $1 \times N$. Thus, an $N \times (N+1)$ MRR crossbar array is needed for the calculation of MVM including negative values in total. This implies that the effect of the offset calculation for obtaining negative values on the computational complexity of MVM can be negligible. Note that an input vector does not include negative values due to the output of the ReLU function. The subtraction of the output values of the MRR crossbar array and ReLU function was executed in a personal computer. We found that optical signals from multiple input ports induce a large noise due to the high coherence of the input signals, which is also discussed in Supplementary Section VI. Thus, the add operation of MVM was also carried out in the computer. We performed ex situ training on a personal computer using 50 instances in the dataset to determine the weights in the ONN shown in Fig. 3(a). Next, we programmed the MRR crossbar array with feedback control to perform MVM according to the trained model. Lastly, we evaluated the ONN based on the MRR crossbar array by testing the classification of 100 instances that were not used in the training. Figures 3(c) and (d) show the results of the classification task using a conventional computer and an ONN, respectively. The implemented ONN correctly identified 93/100 instances, while the conventional computer identifies 97/100 instances. This result shows the feasibility of the ONN using the MRR crossbar array.

**Simulation of on-chip back propagation.** We also performed a simulation of on-chip backpropagation using the ONN based on the MRR crossbar array. In this simulation, we demonstrated the training of a three-layer neural network for the classification of the Iris dataset, which is the same as that shown in Fig. 3(a). Figure 4(a) shows a schematic of the simulation, where $\boldsymbol{x}^{(i)}$, $\boldsymbol{W}^{(i)}$, and $\boldsymbol{\delta}^{(i)}$ ($i = 1, 2$) are the input vector, weight matrix, and auxiliary error vector for back propagation in the $i$-th layer of the ONN, respectively. $\boldsymbol{y}$ represents the output vector of the ONN. The ReLU function $f^{(1)}$ was used as the activation function between the input layer and the hidden layer. Also, the Softmax function $f^{(2)}$ was used as the activation function at the output layer and cross entropy was employed as the loss function for the training $L$. Then, by using backpropagation, stochastic gradient decent was carried out to update the parameters of the neural network, which correspond to phase shifts of the phase shifters in the MRR crossbar array. In this simulation, MVMs with Forward and Backward signals of the $4 \times 4$ MRR crossbar array shown in Figs. 2(a) and (e) were numerically obtained considering the resonance characteristic of an MRR and its crosstalk among the PICs. Note that the transmission of the MRR does not change completely linearly around the resonance state, which may inhibit on-chip backpropagation. The quality factor (Q-factor) of an MRR was assumed to be approximately 9000. In our previous experiment, the MRR crossbar array was used twice to calculate negative values of the weight matrix. On the other hand, the negative values of the input vector are also required to perform on-chip backpropagation. In this case, the calculation related to the offset value is required not only for the weight matrix, but also for the input vector. Thus, the MRR crossbar array was used four times for an MVM, the details of which are discussed in Supplementary Section VII. Figure 4(b) shows the result of the simulation of on-chip backpropagation. All attributes of the Iris dataset were used for training in one training round. The correct rate was defined as the prediction accuracy of the training data. As shown in the figure, the correct rate of the ONN reached 95%. Thus, we have numerically revealed that the proposed on-chip backpropagation using the MRR crossbar array can be used to train ONNs regardless of the nonlinear transmission characteristics of an MRR.

**Benchmark.** Finally, we numerically analyzed the scalability of the circuit size, computation speed, and power consumption of the MRR crossbar array. The details of the analysis are also discussed in the Supplementary Section VII. First, we evaluated the relationship between the circuit size of the MRR crossbar array and the required Q-factor of an MRR to establish the proposed crossbar array. Since the number of channels in the WDM input signals increases with the circuit size, high-Q MRRs are required to distinguish the channels. The permissible overlapping of spectra is evaluated from the supposed computational accuracy that was defined to be 8 bits. Figure 5(a) shows the relationship between the circuit size and the required Q-factor of an MRR. To calculate the Q-factor, we supposed an MRR based on a Si strip waveguide with a radius of 10 μm at a wavelength of 1.55 μm. In this case, we can define the group index and the length of the ring waveguide as 4.2 and 62.8 μm, respectively. The Q-factor of an MRR can be calculated from the ratio of the full width at half maximum (FWHM) to the resonance wavelength. Since the realizable circuit size changes almost proportionally to the FWHM, the Q-factor increases linearly in the figure. When the circuit size is 100, the Q-factor of a MRR must be $2\times10^5$. Since a Si MRR with a Q-factor of $2.2\times10^7$ has been demonstrated[27], a 100 × 100 MRR crossbar array can be fabricated. A microdisk resonator is another candidate for further scaling down the circuit size owing to its high-Q resonance[28,29,30]. Next, we evaluated the computational speed and power consumption of the MRR crossbar array. Supposing that the PIC is controlled with a conventional electrical hardware, the clock frequency for an MVM was assumed to be 3 GHz. When the circuit size of the MRR crossbar array is 100 × 100, the computation speed is expected to be approximately 60 TOPS. To evaluate the power consumption, we assumed the power required for sending and receiving an optical signal as 6.6 pJ/clock, which includes the power of an analog-to-digital converter[31], a digital-to-analog converter[32], and other systems of optical interconnection such as lasers, transimpedance amplifiers, PDs, and so on. In addition, the power of phase shifters should be taken into account. When TO phase shifters, which require 20 mW for a π phase shift, are used to control the MRR crossbar array, the total power consumption is 4.5 W to operate a 100 × 100 MRR crossbar array with the supposed clock frequency. On the other hand, when we employ III-V/Si hybrid metal-oxide-semiconductor (MOS) optical phase shifters[33], their power consumption, which is dominated by only the gate leakage current of its MOS capacitor, is negligible[34], resulting in 2 W power consumption for a 100 × 100 MRR crossbar array. By using a 30-nm-thickness III-V membrane, We can fabricate MRR switches based on a III-V/Si hybrid MOS optical phase shifter[35]. We compare the estimated computation efficiency of the MRR crossbar array with that of 8-bit state-of-the-art electronic hardware for a deep learning accelerator as shown in Fig. 5(b). The red and blue dashed lines show the estimated power consumption and computation speed at various circuit sizes of the MRR crossbar array with III-V/Si hybrid MOS optical and TO phase shifters, respectively. The cases of 100-channel circuit size are highlighted with the red and blue points. The black points show the benchmark of the existing and projected hardware based on application-specific integrated circuits (ASICs), graphics processing units (GPUs), field-programmable gate arrays (FPGAs), and central processing units (CPUs)[3]. The purple dashed lines show computation efficiencies of 1 TOPS/W and 10 TOPS/W. Regardless of the type of phase shifter, the estimated computation efficiency of the MRR crossbar array exceeds 10 TOPS/W, which is much higher than that of the existing electronic hardware. Also, in the case of the hybrid MOS optical phase shifter, the computation efficiency is predicted to be 30 TOPS/W. This result shows that the MRR crossbar array has the potential to greatly improve the computation efficiency of inference. Moreover, we evaluated the computation time of training when using the ONN. Figure 5(c) shows the calculated computation time when using the MRR crossbar array to update the parameters of the three-layer ONN once, supposing a 3 GHz clock frequency for an MVM. In the case of brute force gradient calculation discussed in Ref. 8, the calculation time increases proportionally to the number of parameters in the ONN, whereas our proposed method using on-chip backpropagation requires only three optical-domain MVMs regardless of the circuit size, resulting in 2000 times shorter computation time than the brute force gradient method when the circuit size is 100 × 100. Thus, the MRR crossbar array has a great potential to accelerate learning in addition to inference.

**Discussion**

In this paper, we demonstrated a proof-of-concept MRR crossbar array for both inference and training tasks. By using a fully integrated prototype chip, we successfully implemented ONN based on the MRR crossbar array and demonstrated the classification task of the IRIS dataset with a prediction accuracy of 93 %. Also, we proposed on-chip backpropagation using the MRR crossbar array and demonstrated its feasibility. Furthermore, we revealed that our proposed PIC has the potential to perform with a computational efficiency of 30 TOPS/W. The advantages of our ONN based on the MRR crossbar array compared with ONNs based on the other Si programmable PICs are summarized as follows. (a) MRR-based PICs can reduce circuit size and power consumption compared with MZI-based PICs. Previous studies revealed that the MRR crossbar array can reduce its circuit size to be 1/36 that of a the MZI-based PIC[20]. Also, since an MRR can realize large switching with a small phase shift, the power consumption required for phase shifters of the MRR crossbar array is also smaller than that of meshed-MZI. (b) It is simple to program the MRR crossbar array to perform MVM for the inference task. In the case of meshed-MZI, complex calibration methods are needed to use it for MVM. On the other hand, since the MRR crossbar array works with WDM optical signals, it is easier to control its output for MVM owing to feedback calibration. (c) The MRR crossbar array has the potential to accelerate the training task. In addition to the feedback calibration, the MRR crossbar array can potentially perform on-chip backpropagation for transpose matrix operation, which improves the training task of deep learning. (d) Negative values can be calculated by introducing offset calculation. In the case of meshed-MZI, a negative value is expressed with the phase of the optical signal, which requires homodyne detection to measure the value. In Ref. 10, on-chip homodyne detection is introduced to increase the stability and credibility of the measurement, which however requires multiple and complex detection steps. Also, the weight bank uses add/drop characteristics of an MRR and balanced PDs for the calculation of the negative values, which however generates an inner loop in the PIC and restricts the circuit size. In this paper, the offset value was used to calculate MVM including the negative weight matrix. This method is simple and scalable among these PICs. With these advantages, the MRR crossbar array is promising for building scalable, efficient deep learning accelerator for both inference and training tasks.

**Methods**

**Experimental setup.** A tunable laser diode (TLD) and a multiplexer (MUX) unit were used to generate multi wavelength optical signals at wavelengths $\lambda_1$, $\lambda_2$, $\lambda_3$, and $\lambda_4$ shown in Fig. 1(a). Photocurrents from the Ge PDs were received using photocurrent amplifiers and analog-to-digital converters (A/Ds). All phase shifters were electrically controlled by a 40-ch multi output direct current (DC) source. The DC source was driven in the constant-current (CC) mode to reduce noise caused by unstable connections. A personal computer with a 64-bit central processing unit (CPU) was used to adjust current applied to the phase shifters and to read out the output currents of the PIC. The Python framework was also employed to control these instruments. The chip of the MRR crossbar array was mounted on an aluminum-plate-based printed circuit board (PCB). The aluminum-plate-based PCB has a higher thermal conductivity, resulting in a small thermal crosstalk on the PIC. For electrical connections between the PCB and the other measurement devices, a flexible flat cable (FFC) was used to reduce the complexity of electrical connection.

**Acknowledgements**

This work was partly supported by JST CREST Grant Number JPMJCR1907.

**Author contributions**

S. O. contributed to the idea, device design, measurement, simulation, and manuscript preparation. T. K. and S. T. contributed to the discussion and high-level project supervision. M. T. contributed to the idea, discussion, manuscript revision, and high-level project supervision.

**Competing financial interests**

The authors declare no competing financial interests.

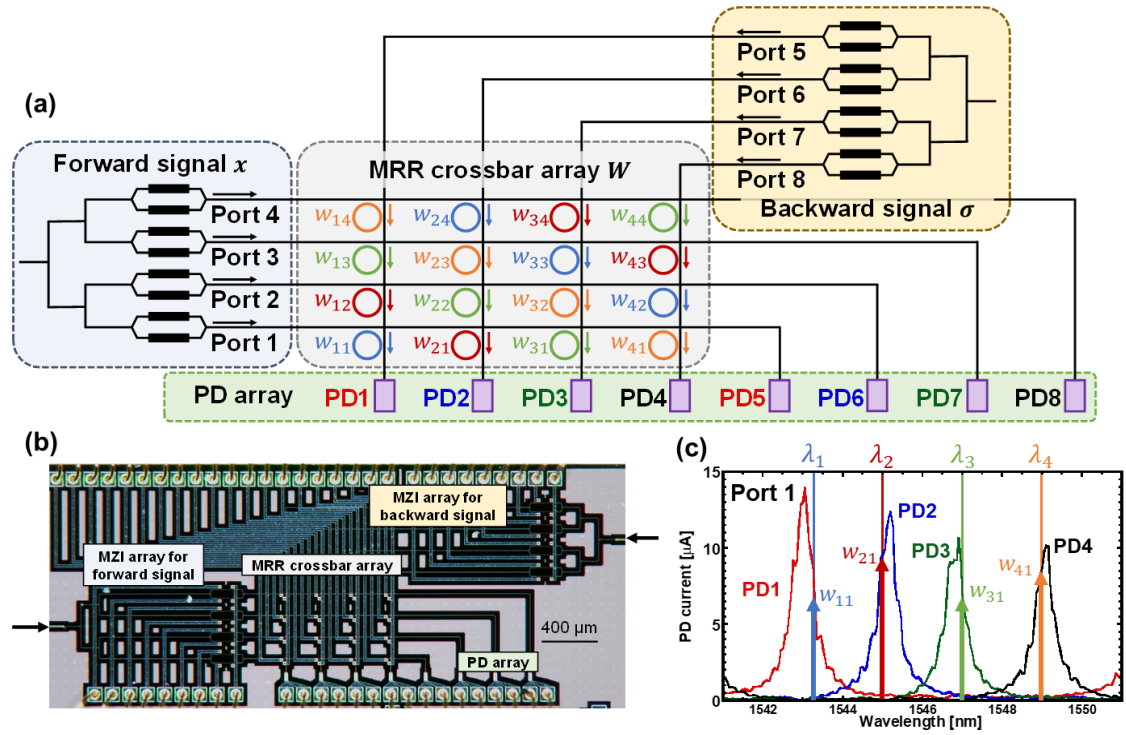

Fig. 1. **Fully integrated MRR crossbar array. a,** Schematic of MRR crossbar array including couplers, MZI switches, and PDs. **b,** Plan-view image of MRR crossbar array corresponding to **a**. **c,** Output spectra of MRR crossbar array when optical signal is injected from Port 1.

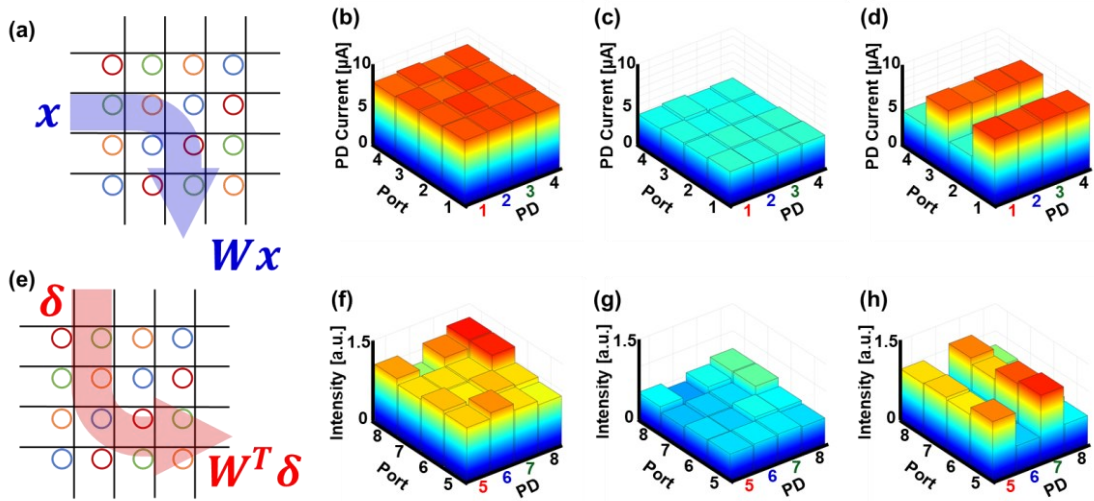

Fig. 2. **Basic operations of MRR crossbar array. a,** Route of forward signal. **b–d,** Output of the MRR crossbar array with forward signal, which was calibrated with feedback control. **e,** Route of backward signal. **f–h,** Output of MRR crossbar array with backward signal after calibration with forward signal. These outputs were corrected with parameters related to input ports and PDs.

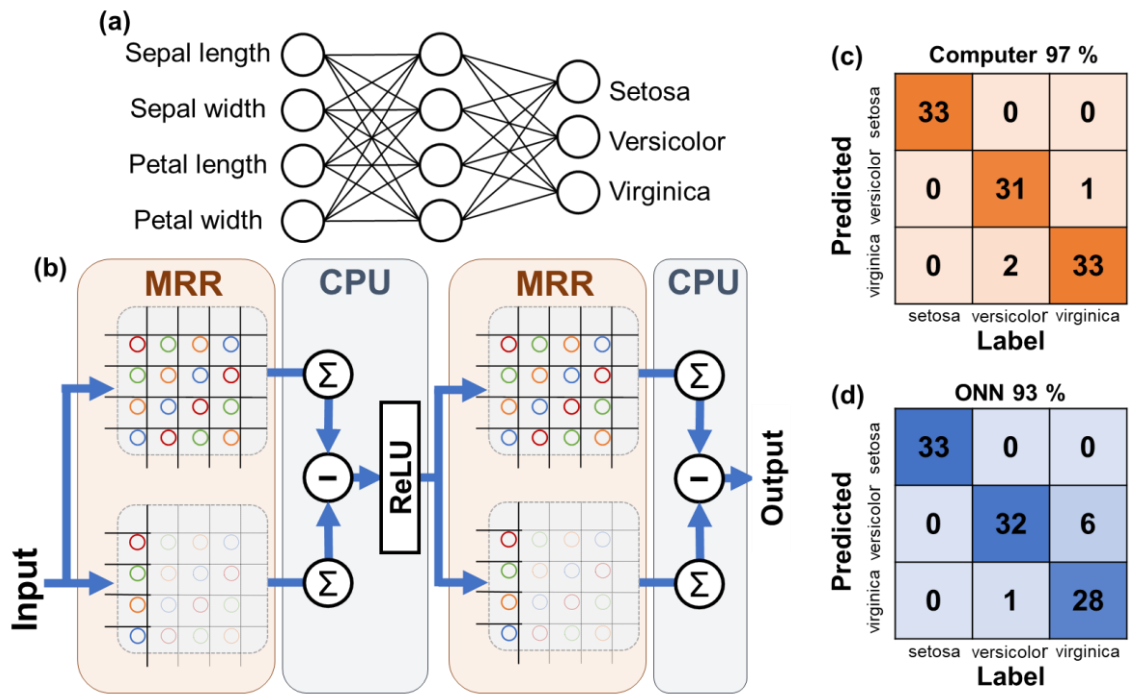

Fig. 3. **Inference task of three species Iris flowers using ONN based on MRR crossbar array. a,** Schematic of neural network for demonstration. **b,** Schematic of implementation of ONN. **c,** Label prediction result of neural network using conventional computer. **d,** Label prediction result of ONN.

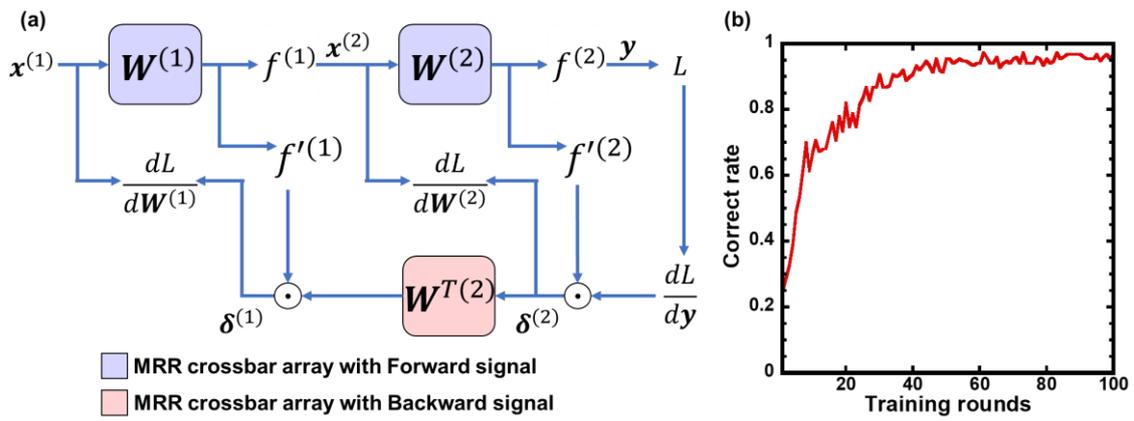

Fig. 4. **Simulation of on-chip backpropagation using ONN based on MRR crossbar array. a,** Schematic of simulation. **b,** Result of training using IRIS dataset.

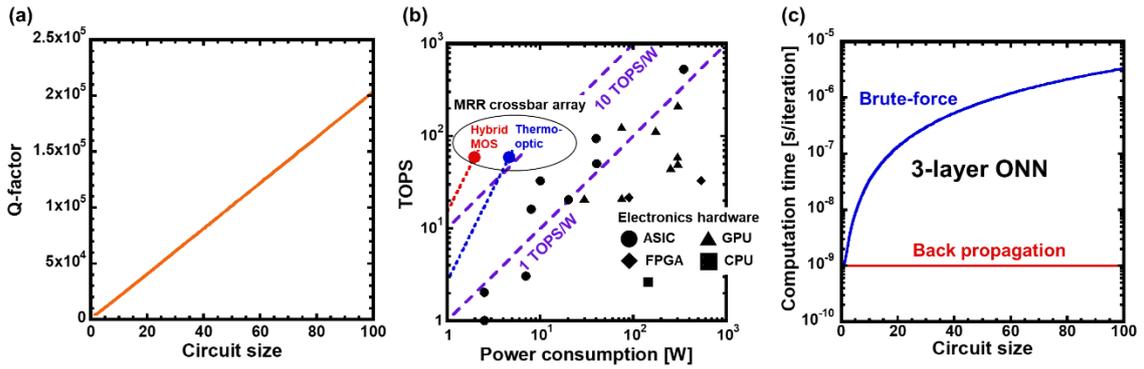

Fig. 5. **Benchmarks of MRR crossbar array. a,** Calculated relationship between circuit size of MRR crossbar array and required Q-factor of MRR. **b,** Calculated computation efficiency of MRR crossbar array and conventional electronics hardware. **c,** Calculated computation time of on-chip backpropagation using MRR crossbar array.

# Supplementary Information

# Si microring resonator crossbar array for on-chip inference and training of optical neural network

Shuhei Ohno[1*], Kasidit Toprasertpong[1], Shinichi Takagi[1], and Mitsuru Takenaka[1**]

[1]The University of Tokyo, 7-3-1 Hongo, Bunkyo-ku, Tokyo 113-8656, Japan

*E-mail: ohno@mosfet.t.u-tokyo.ac.jp

**E-mail: takenaka@mosfet.t.u-tokyo.ac.jp

# I. Characteristics of MRR and PD in MRR crossbar array

Figure S1(a) shows a plan-view photograph of a microring resonator (MRR) with a thermo-optic (TO) phase shifter. A 210 nm × 400 nm Si strip waveguide was used as a single-mode waveguide at a wavelength of 1550 nm. The gap and radius of the MRR were designed to be 10 μm and 200 nm, respectively. A TiN heater was deposited on the ring waveguide. The waveguide crossing consisting of four ellipse waveguides was used at the intersection to suppress its insertion loss and crosstalk[1]. Figure S1(b) shows output spectra of the MRR. These characteristics were evaluated using an integrated Ge photodetector (PD). The free spectral range (FSR) and quality factor (Q-factor) of the MRR were evaluated to be 9 nm and 5200, respectively.

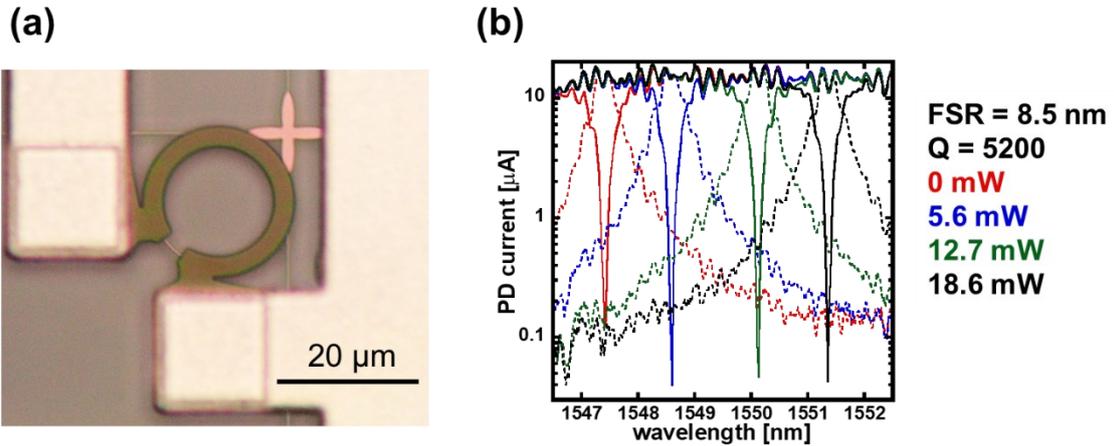

**Fig. S1. (a) Plan-view microscopy image and (b) output spectra of MRR with TO phase shifter.**

Figures S2(a) and (b) show a plan-view photograph and a cross-sectional schematic of the Ge PD integrated with the MRR crossbar array, respectively. The Ge PD consisted of the vertical PIN junction with p-Si, i-Ge, and n-Ge layers. The Ge layer was grown on the tapered Si waveguide. Figure S2(c) shows I–V curves of the PD at various optical powers at a wavelength of 1550 nm. In this study, no bias voltage was applied to the PD to make the measurement setup simple.

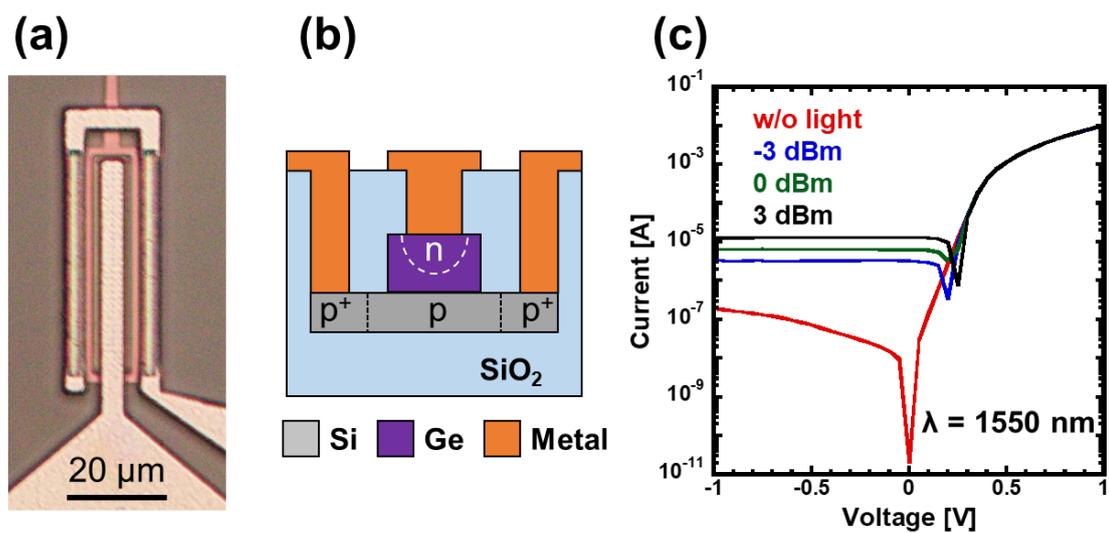

**Figure S2. (a) Plan-view photograph, (b) cross-sectional schematic, and (c) photocurrent of Ge PD.**

## II. Experimental setup of the MRR crossbar array

Figure S3(a) shows a schematic of an experimental setup for the demonstration of an ONN based on the MRR crossbar array. Figure S3(b) shows a picture of the measurement setup. Multiple tunable laser diodes (TLDs) and a multiplexer (MUX) unit were used to generate multi wavelength optical signals at wavelengths $\lambda_1, ... ,\lambda_4$ shown in Fig. 1(c). The photocurrents from the Ge PDs were received using photocurrent amplifiers and analog-to-digital converters (A/Ds). All phase shifters were electrically controlled by a 40-ch multi output direct current (DC) source. The DC source was driven in the constant-current (CC) mode. A personal computer with a 64-bit central processing unit (CPU) was used to control injection currents to the phase shifters and read out the output currents of the photonic integrated circuit (PIC). Figure S3(c) shows the prototype chip of the MRR crossbar array packaged on a printed circuit board (PCB). This PCB has an aluminum plate for its base. Compared with well-known PCBs such as glass composite PCBs, this PCB has a higher thermal conductivity, resulting in a small thermal crosstalk throughout the PIC. For electrical connections between the PCB and the other measurement devices, a flexible flat cable (FFC) was used to reduce the complexity of electrical connection. Figure S3(d) shows the connection between the cleaved fiber and the spot-size converter (SSC) of the PIC. By connecting the fiber without any space, we can eliminate vibrations of the fiber caused by the fiction force between them, reducing the measurement noise.

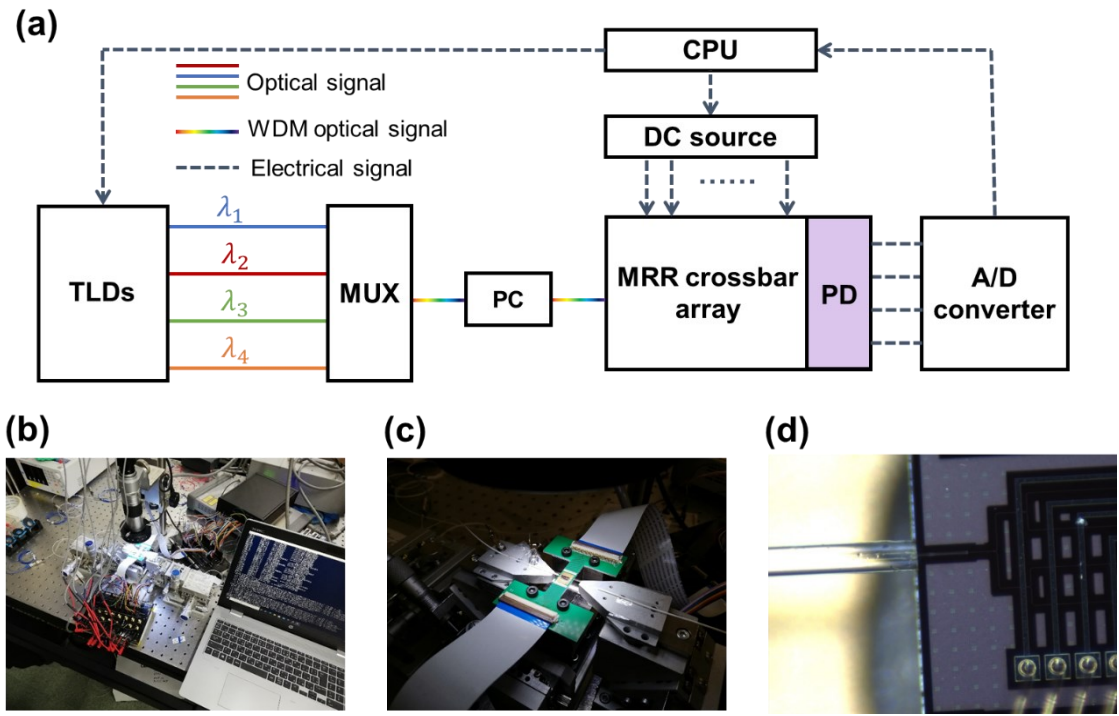

**Fig. S3.** (a) Schematic and (b)–(d) photographs of experimental setup of MRR crossbar array.

## III. Characterization of MRR crossbar array

Figure S4(a) shows the output spectra of the MRR crossbar array measured by the photocurrents of PDs 1–4 when an optical signal was injected in the forward direction. These spectra contain drop characteristics of the 16 MRRs in the PIC. Before the measurement of the spectra, we roughly tuned the drop peak of the MRRs with currents applied to the phase shifters to fit the assigned wavelengths of the wavelength division multiplexing (WDM) optical signal. The assigned wavelengths were defined according to the scheme discussed in Supplementary Section VII to expand the wavelength margin between the neighbor channels of the MRRs. The four lines in the figure show the wavelengths of 4-ch WDM signals used for matrix-vector multiplication (MVM) in Fig. 2. The sixteen arrows in the figure correspond to the weight elements of the matrix for MVM shown in Fig. 1(a). We observed noise in the spectra, which can be attributed to the reflection at the MMI couplers in the MZIs. Figure. S4(b) shows the output spectra of the MRR crossbar array measured by PDs 5–8 when the WDM optical signal was injected in the backward direction. It can be seen that the transpose output was obtained by injecting the same currents to the phase shifters as in the forward direction.

In the measurements shown in Fig. S4, the optical signal was injected to a single input port to the MRR crossbar array by tuning MZI switches. To evaluate the feasibility of the add operation in MVM, we evaluated the output spectra when the optical signals were injected to the multiple input ports simultaneously, as shown in Fig. S5. The red line shows the output spectra when the optical signal was injected to only Port 1, while the blue, green, black lines show those when the optical signals were additionally injected to Ports 2, 3, and 4, respectively. Focusing on the assigned wavelengths $\lambda_1, ..., \lambda_4$, the outputs show the noise due to the crosstalk, which affects the accuracy of MVM. This noise occurred not only at the peak of MRR but also at all wavelengths. Therefore, the origin of the noise might be not only the crosstalk between the multiple MRRs in the PIC. To investigate this origin, the output spectra of the MRR crossbar array must be measured with a fiber array to eliminate the effect of MMI couplers and MZI switches, which will be discussed in Supplementary Section VI.

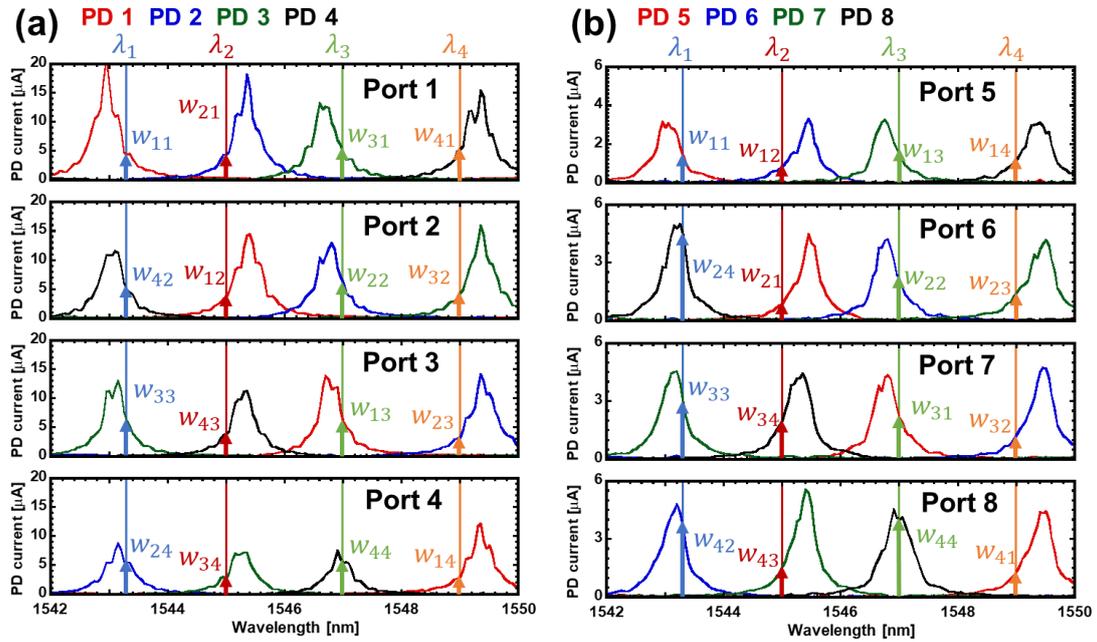

**Fig. S4.** Output spectra of MRR crossbar array with (a) Forward and (b) Backward signals.

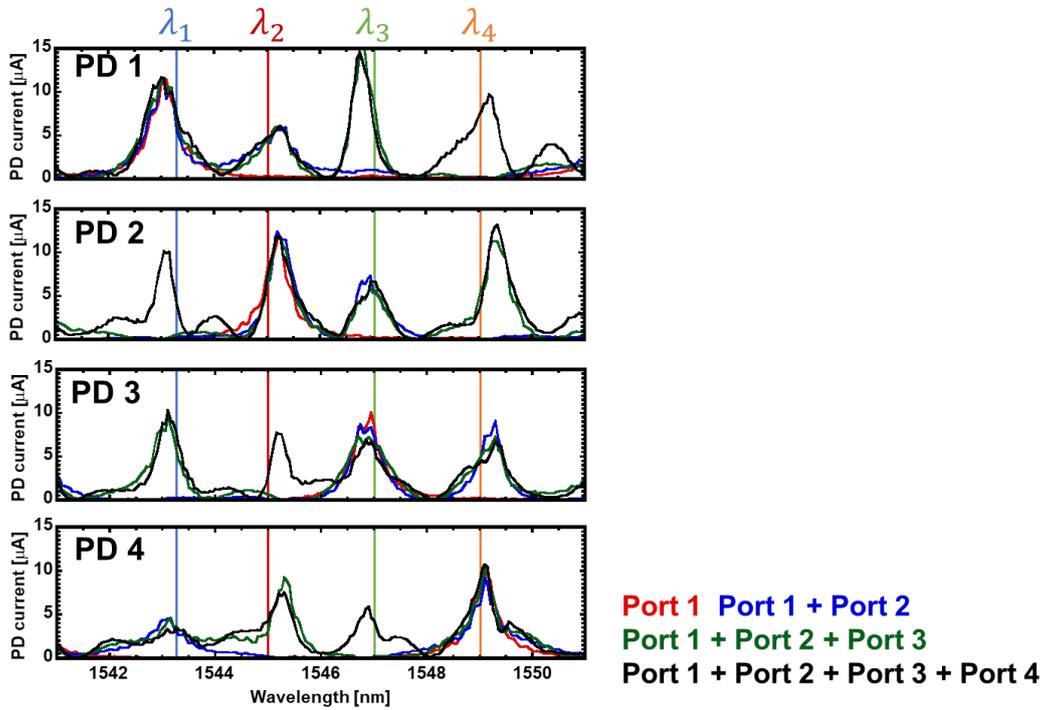

**Fig. S5.** Output spectra of MRR crossbar array with multiple input signals.

## IV. Calibration for transpose matrix operation

Figures S6 (a)–(c) show the outputs of the MRR crossbar array with the Backward signal. Before the measurement, the currents injected to the phase shifters of the PIC were adjusted to obtain the outputs shown in Figs. 2(b)–(d) with the feedback calibration[2]. Figures S6 (a)–(c) show the outputs of the transpose matrix operation with the Backward signal, which correspond to those of the matrix operation with the Forward signal shown in Figs. 2(b)–(d), respectively. However, there is the mismatch between the outputs in Fig. S6 and the transpose matrices in Fig. 2(b)–(d).

Figure S7(a) shows the main origin of the mismatch between the outputs in Fig. 2(b) and the transpose matrices in Fig. S6(a). In this case, the phase shifters of the MRRs in the PIC were programmed to obtain the same transmission for the Forward signal. However, in the case of the transpose operation, there is variation in output power due to the optical path difference. Thus, we needed the calibration for the Backward signal to obtain the transpose output by applying the constant multiplication with eight parameters corresponding to each input port (A5, ..., A8) and PD (B5, ..., B8) as shown in Fig. S7(b). Figures 2(f)–(h) show the results after calibrating the results in Fig. S6. Note that we can design an MRR crossbar array to have an equivalent path length in any combinations between the input and output ports.

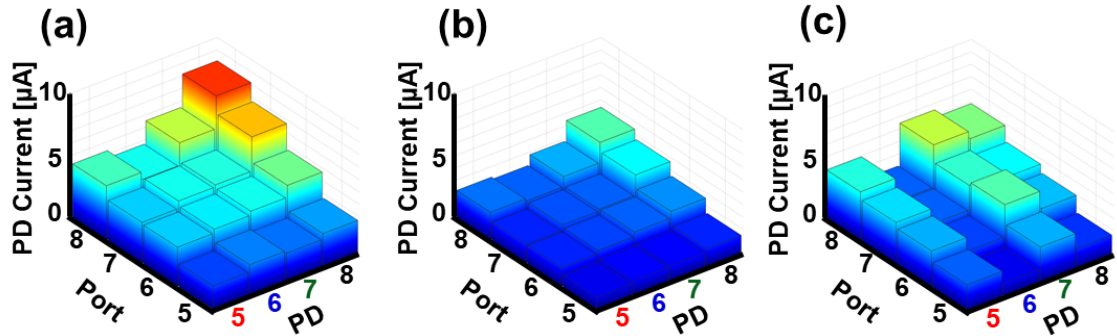

**Fig. S6. (a)–(c) Outputs of MRR crossbar array with Backward signal.**

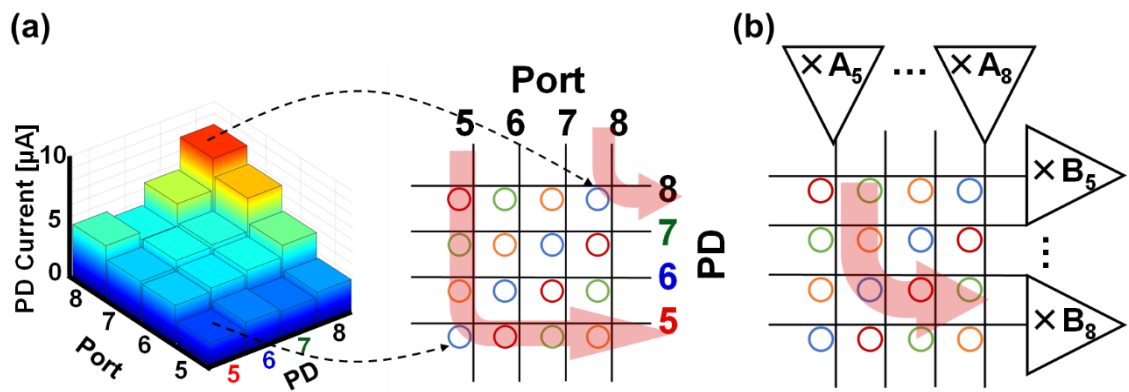

**Fig. S7. (a)** Schematic of origin of error in Fig. S6, showing longest and shortest paths in backward propagation. **(b)** Schematic of calibration for backward calculation.

## V. MVM including negative values

To obtain the results in Figs. 3 and 4, MVM including negative values was required, while the MRR crossbar array cannot simply calculate negative values. Thus, we introduced an offset matrix $W_{\text{offset}}$ to perform MVM including negative values. Figure S8 shows a schematic of MVM using the MRR crossbar array including negative matrix elements using a simple example. First, we add an offset matrix to an original weight matrix $W_{\text{weight}}$. The matrix after this operation $W'_{\text{weight}}$ is expressed as

$$W'_{\text{weight}} = \frac{1}{2}W_{\text{weight}} + W_{\text{offset}} = \frac{1}{2}W_{\text{weight}} + \frac{1}{2}\begin{pmatrix} 1 & \cdots & 1 \\ \vdots & \ddots & \vdots \\ 1 & \cdots & 1 \end{pmatrix},$$

where $W_{\text{offset}}$ is a matrix, whose elements have the same positive value. The offset value is determined so that all elements of $W'_{\text{weight}}$ are non-negative values. Note that the range of the elements of $W_{\text{weight}}$ is fixed to be $[-1, 1]$ after normalization. Thus, the range of the elements of $W'_{\text{weight}}$ is $[0, 1]$. Next, we perform MVM for both $W'_{\text{weight}}$ and $W_{\text{offset}}$ with the input vector $x$, resulting in $W'_{\text{weight}}x$ and $W_{\text{offset}}x$, respectively. Since all elements of $W_{\text{offset}}$ are the same, the size of the MRR crossbar array for the calculation of $W_{\text{offset}}x$ can be suppressed, as shown in Fig. S8. Finally, $W_{\text{weight}}x$ can be obtained as

$$W_{\text{weight}}x = 2(W'_{\text{weight}}x - W_{\text{offset}}x).$$

We also consider the computational complexity of this calculation. If the size of $W_{\text{weight}}$ is $N \times N$, a $N \times N$ MRR crossbar array is required for MVM of $W'_{\text{weight}}x$. On the other hand, only a $1 \times N$ MRR crossbar array is needed for MVM of $W_{\text{offset}}x$ since all elements of $W_{\text{offset}}$ are the same as mentioned above. In total, a $N \times (N + 1)$ MRR crossbar array is needed for the calculation of MVM including negative values. Since an MRR required for calculating $W_{\text{offset}}x$ does not change proportionally to the square of $N$, the effect of $W_{\text{offset}}$ on the computational complexity can be negligible when $N$ is large.

Next, we discuss MVM including both a negative matrix and a negative input vector. In particular, a negative input vector is required for on-chip training. Note that MVM including a negative input vector was not used in the demonstration of the inference task shown in Fig. 3 because we used the ReLU function as the activation function. Figure S9 shows its schematic. To deal with a negative input vector, the method using offset values described previously is applicable. In total, we need to operate MRR crossbar array 4 times for a MVM including negative value.

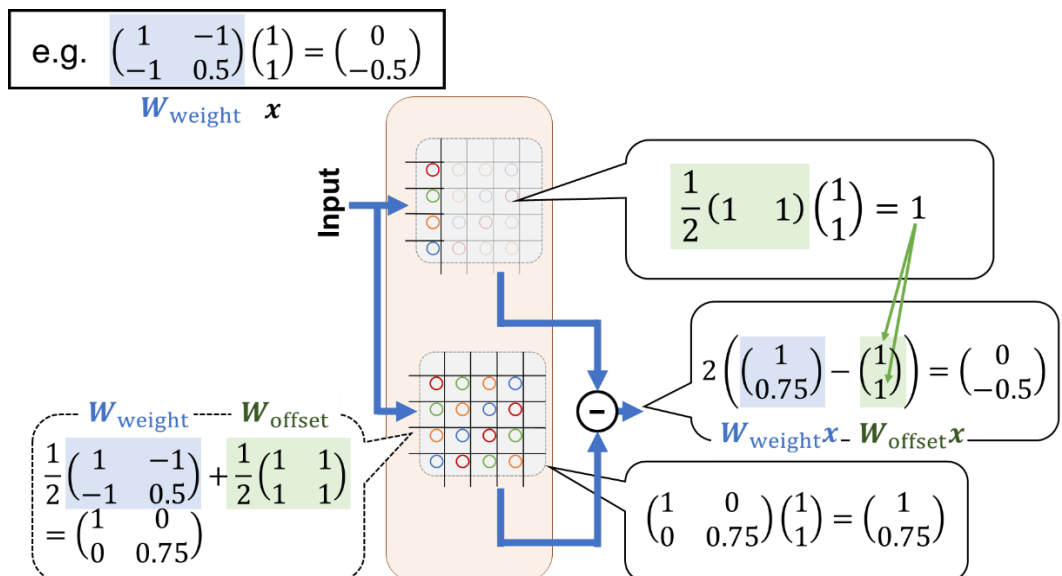

**Fig. S8. Schematic of MVM including negative matrix elements.**

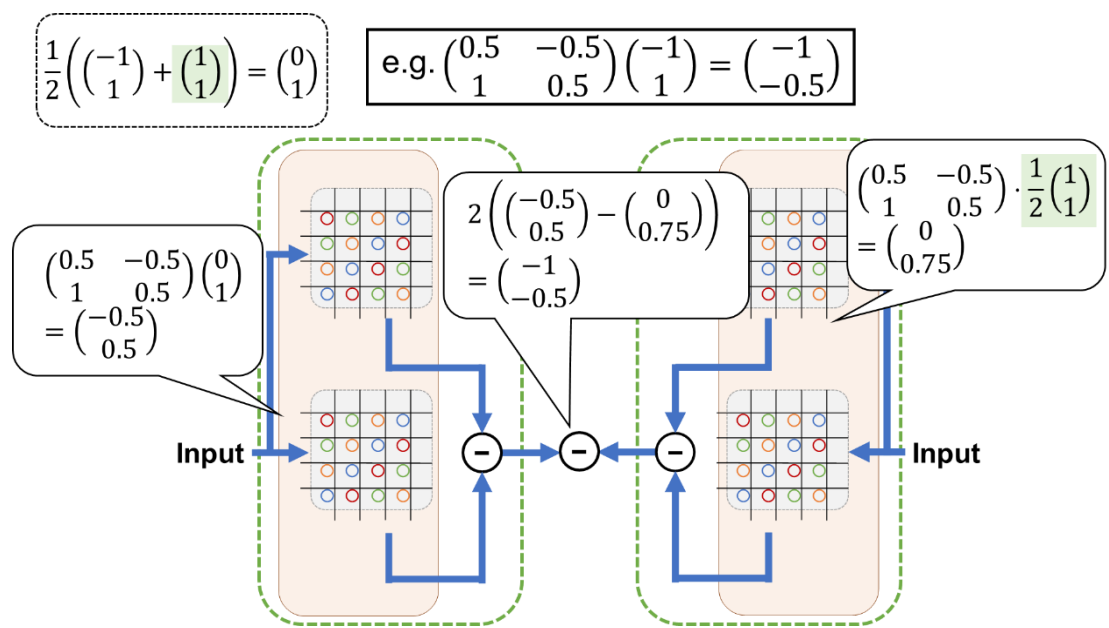

**Fig. S9. Schematic of MVM including both negative input vector and negative matrix elements.**

## VI. Measurement of MRR crossbar array using fiber array

As shown in Fig. S5, we could not perform an accurate add operation when the WDM signals were simultaneously input to the multiple ports of the MRR crossbar array. To investigate the origin of this issue, we measured another prototype of the MRR crossbar array without the splitters and MZI modulators for input vector generation. For multiple inputs, we used an optical fiber array. Figure S10 shows the plan-view microscopy image and schematic of the MRR crossbar array for this measurement. This PIC consists of nine MRRs for 3 × 3 MVM. To reduce the coupling loss between the fiber array and Si waveguides, we prepared spot-size converters (SSCs) for the input ports. The two outer SSCs are directly connected by a straight waveguide, which is used for fiber array alignment.

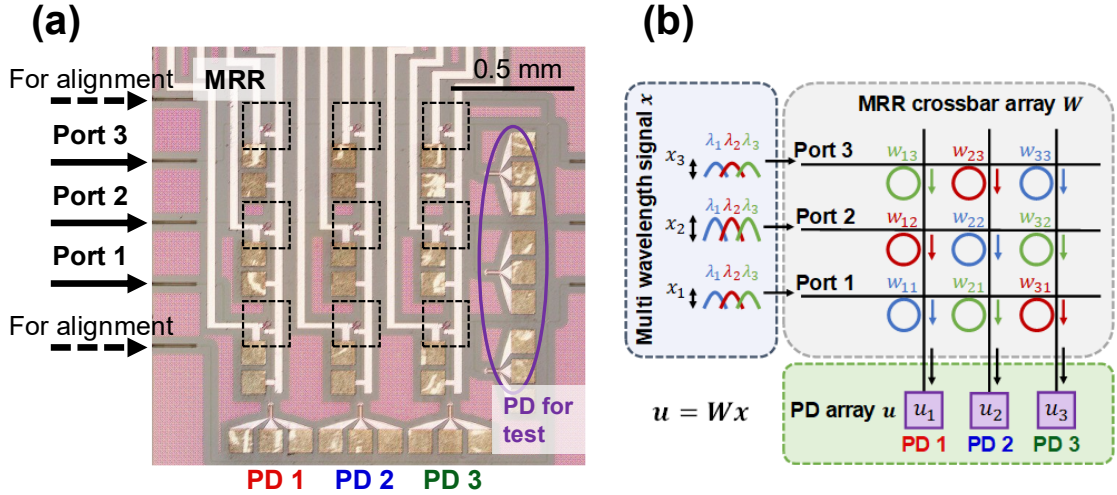

**Fig. S10. (a) Plan-view microscopy image and (b) schematic of MRR crossbar array for measurement using fiber array.**

Figure S11 shows a schematic of the measurement setup. A TLD was used for optical input signals of the PIC. The optical input signals were divided using multiple fiber couplers. The divided signals were injected to the three input ports of the MRR and the alignment port for fiber alignment. Another alignment port was connected to a power meter via the fiber array to monitor the transmission of the optical signal. The fiber array was efficiently connected to the multiple input ports by maximizing the intensity. The inset in Fig. S11 shows a photograph of the measurement setup.

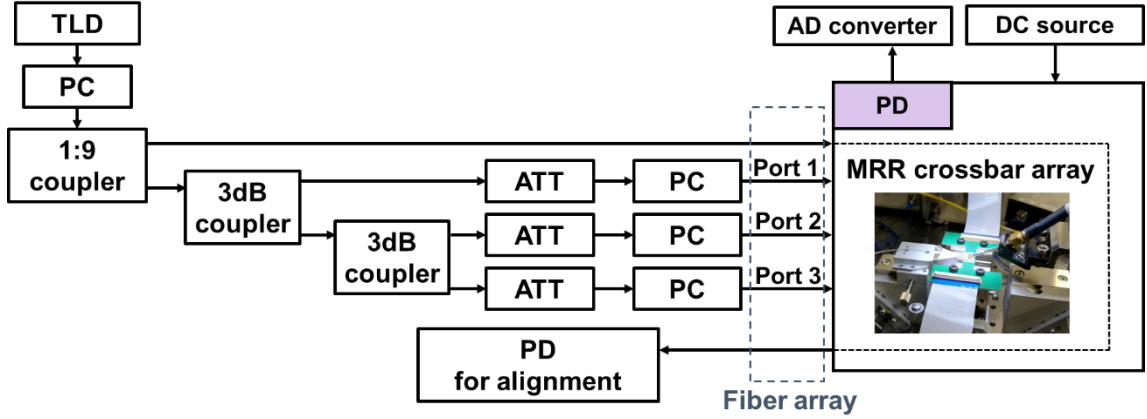

**Fig. S11. Schematic of experimental setup for demonstration of MRR crossbar array with fiber array.**

We measured the wavelength characteristics of the MRR crossbar array. Before the measurement, the phase shifters were calibrated for MVM. First, we examined the output spectra when the optical signal was injected to a single input port. Figure S12(a) shows the output spectra of the MRR. Since the MMI coupler and MZI switches are eliminated from the PIC, compared with Fig. S4, the wavelength characteristics are clear. Next, we measured the output spectra by injecting the optical signal to multiple input ports, as shown in Fig. S12(b). As shown by the blue and green lines in the figure, when the optical signal is injected to the multiple input ports, a small noise is found in the output spectra. Compared with Fig. S5, the spectra are clear, although noise still exists, which disturbs add operations in MVM. Thus, the origin of the noise is not only the MMI and MZI in the PIC, but also undesired interferences inside the PIC. To observe the impact of coherence crosstalk, we prepared two different TLDs for generating multiple optical signals instead of the 3 dB couplers. Figure S13(a) shows a schematic of the measurement setup for the optical input. The optical signals from TLD1 and TLD2 were injected to Ports 1 and 2 via the fiber array, respectively. In the measurement, the wavelengths of optical signals from TLD1 and 2 were swept at the same time. Figure S13(b) shows the output spectra obtained. Compared with Fig. S12(b), the noise in the spectra was eliminated. This result reveals that the noise in Figs. S5 and S12 was caused by undesired coherent crosstalk. We can mitigate the impact of coherence crosstalk by optimizing the design and fabrication accuracy. As shown in Fig. S13, the preparation of multiple laser sources and time-multiplexed calculation are also effective.

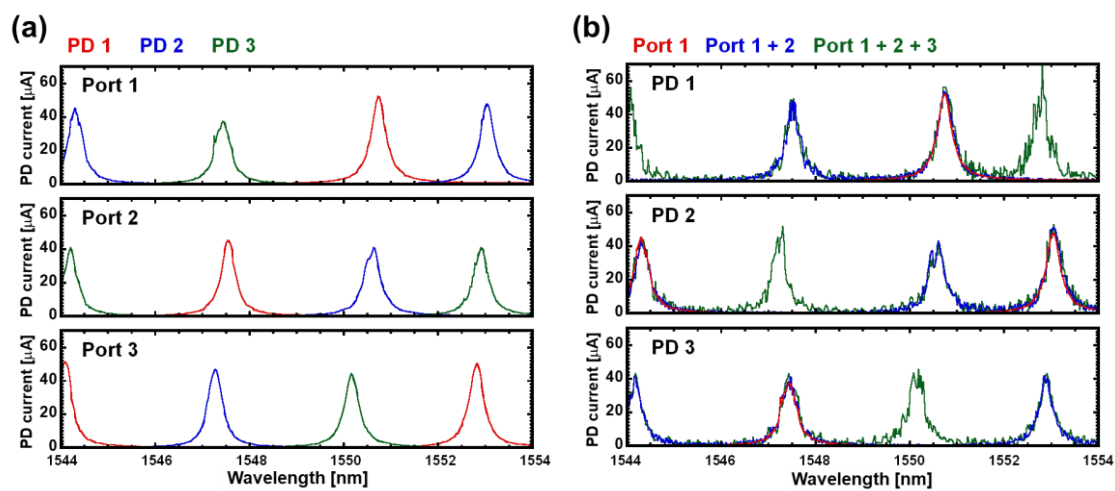

**Fig. S12. Output spectra of MRR crossbar array with (a) single input and (b) multiple inputs.**

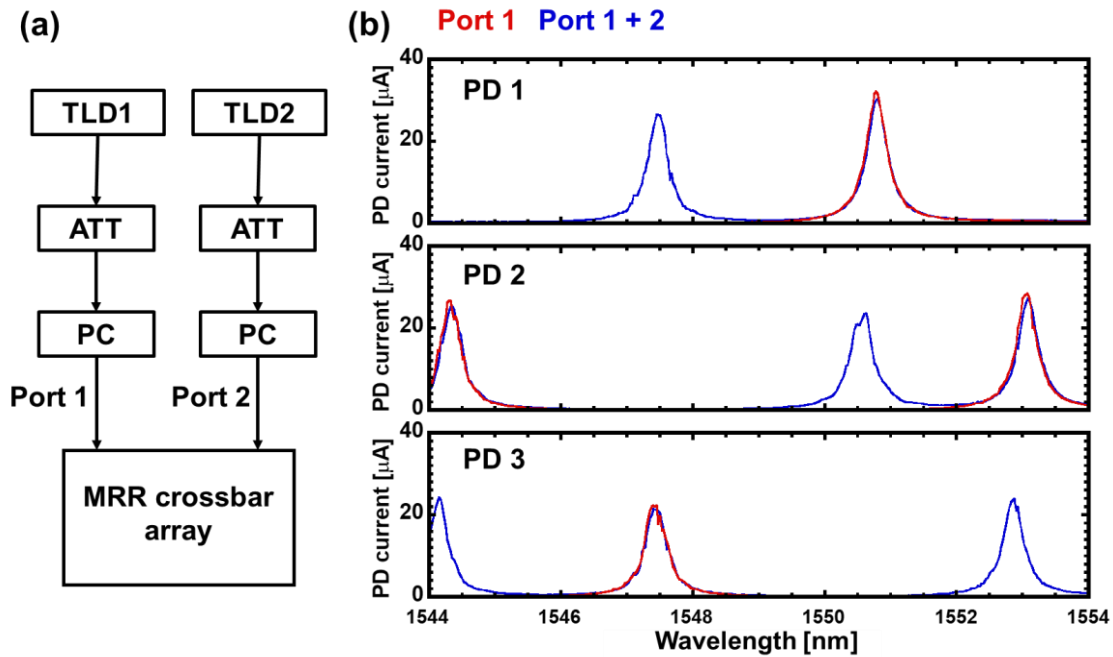

**Fig. S13. (a) Schematic of experimental setup with multiple TLDs and (b) obtained output spectra.**

## VII. Performance analysis of MRR crossbar array

In this section, we discuss the feasible circuit size of the MRR crossbar array. Also, we evaluate computation speed and power consumption by numerical analysis to compare with other state-of-the-art hardware for deep learning.

Before the analysis, we review the characteristics of an MRR[3]. Figure S14(a) shows a schematic of an MRR, which has the PASS and DROP bus waveguides. Then, the transmission of the PASS and DROP ports in the MRR $T_\mathrm{p}$ and $T_\mathrm{d}$ can be expressed as

$$T_\mathrm{p} = \frac{r_2^2 a^2 - 2r_1 r_2 a \cos\phi + r_1^2}{1 - 2r_1 r_2 a \cos\phi + (r_1 r_2 a)^2}, \qquad (1)$$

$$T_\mathrm{d} = \frac{(1-r_1^2)(1-r_2^2)a}{1 - 2r_1 r_2 a \cos\phi + (r_1 r_2 a)^2}, \qquad (2)$$

where $r_i\ (i=1,2)$ and $a$ are the self-coupling coefficient of the coupling regions and the single-pass amplitude transmission of the ring waveguide, respectively. $\phi$ corresponds to the phase shift at the ring waveguide. $\phi$ is also expressed as

$$\phi = \frac{n_\mathrm{eff}\lambda}{L}, \qquad (3)$$

where $n_\mathrm{eff}$ and $L$ are the effective refractive index and length of the ring waveguide, respectively. $\lambda$ corresponds to the wavelength of the optical signal. When $\lambda$ is resonance wavelength, $\phi = 2m\pi\ (m=0,1,2,\ldots)$. By using an optical phase shifter, we can tune $n_\mathrm{eff}$, which can control resonance wavelength. Figure S14(b) shows the calculated output spectra of an MRR obtained by (1) and (2). The spectral range of an MRR can be determined by the quality factor (Q-factor), which can be expressed as

$$Q = \frac{\pi n_\mathrm{g} L \sqrt{r_1 r_2 a}}{\lambda_\mathrm{res}(1 - r_1 r_2 a)}, \qquad (4)$$

where $n_\mathrm{g}$ and $\lambda_\mathrm{res}$ are the group index of the ring waveguide and the resonance wavelength of the MRR, respectively. Since the Q-factor is the ratio of the confined optical intensity to the decayed optical intensity in the resonator, we can define the photon lifetime $\tau_\mathrm{p}$ as

$$\tau_\mathrm{p} = \frac{\lambda_\mathrm{res} Q}{2\pi c}, \qquad (5)$$

where $c$ is the speed of light. $\tau_\mathrm{p}$ is associated with the bandwidth of the MRR.

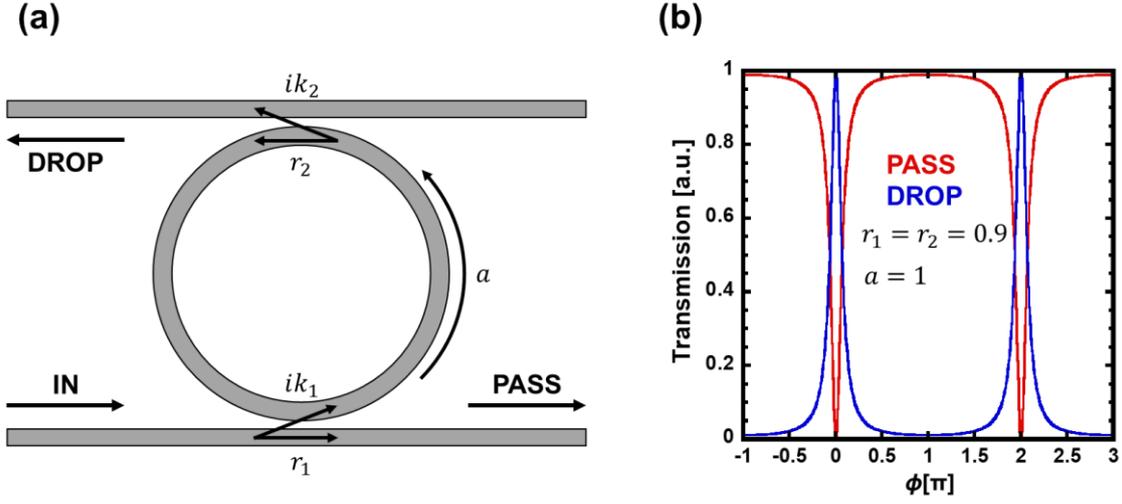

**Fig. S14. (a) Schematic and (b) output spectra of MRR.**

First, we consider the circuit size of the MRR crossbar array $N$, which corresponds to MVM using the $N \times N$ matrix. In this case, the WDM signal consists of $N$ different wavelengths. Figure S15 shows the output spectra of four different MRRs required for the $4 \times 4$ MRR crossbar array as functions of $\phi$, which can be converted to the function of $\lambda$ by (3). The four vertical lines in the figure correspond to different channels of the WDM signal, which are named as Ch. 1–4. 4 colors of the peak correspond to different output spectra of MRRs. The solid line of each drop peak of MRR shows the output used for the multiplication. In the figure, the spectra are located such that the transmission of the corresponding channels of the signal is maximized. To reduce the transmission of each MRR for the MVM of the PIC, their wavelength peaks are moved toward the directions represented by the arrows in the figure according to the phase shift of their phase shifter. To use the MRR crossbar array appropriately, multiple spectra of the MRRs must not overlap at a single channel of the WDM signal. On the other hand, by defining the multiplication region on different sides with neighboring drop peaks as mentioned above, every other distance between adjacent wavelengths of the WDM signal (the distance between Ch. 2 and Ch. 3 in Fig. S15) can be halved compared with others (the distance between Ch. 1 and Ch. 2, and that between Ch. 3 and Ch. 4 in Fig. S15) while satisfying the constraint of the spectrum overlap. In this case, the drop peak of each MRR can be placed in the closest way. Considering that an MRR has $T_\mathrm{d}(\Delta\phi/2)$ noise from its adjacent wavelength channel, $N$ can be calculated as

$$N = \frac{8\pi}{3\Delta\phi}. \quad (6)$$

Then, $\Delta\phi$ can be defined as the width of the drop peak. Considering the $b$-bit computation precision of the PIC, the required condition of $\Delta\phi$ is

$$\frac{T_d(\Delta\phi/2)}{T_d(0)} < 2^{-b}. \quad (7)$$

Using (2), (4), (6), and (7), we calculated the relationship between $N$ and the required Q-factor of the MRR shown in Fig. 5(a). We suppose a computation precision of 8 bits to compare with the existing state-of-the-art hardware for the inference task of deep learning.

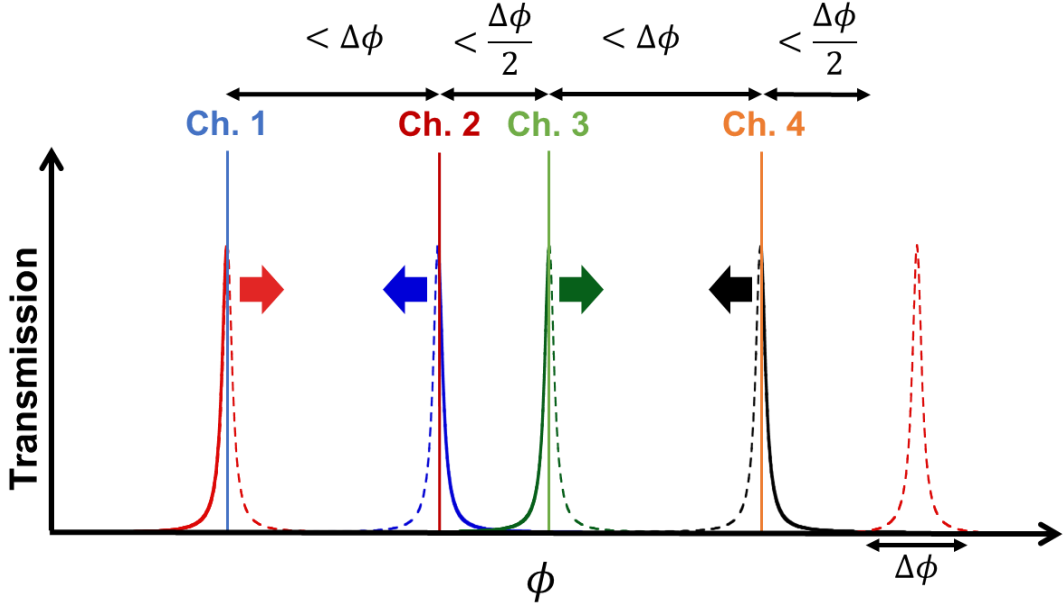

**Fig. S15. Output spectra of MRRs to obtain large-channel MRR crossbar array.**

Next, we estimate the calculation speed and power consumption of the MRR crossbar array at various $N$ values. The computation speed can be calculated as

$$TOPS = 2N^2 f_{CLK} \times 10^{-12}, \quad (8)$$

where $f_{CLK}$ is the clock frequency of the MRR crossbar array and 2 in the equation represents the pair of add and multiply operations. $TOPS$ stands for "Tera Operations Per Second," which is used to evaluate the computation speed of hardware for deep learning. In most cases, a Si programmable PIC requires a host interface based on the electronics platform. Thus, we consider the case of $f_{CLK} = 3$ [GHz], which is sufficiently slower compared with the bandwidth of the MRR calculated from (5). Figure S16(a) shows the relationship between $N$ and $TOPS$ calculated from (8).

The power consumption of the MRR crossbar array per single MVM, $E_{MVM}$, can be calculated as

$$E_{MVM} = f_{CLK} E_{signal} N + \frac{\Delta \phi}{2\pi} N^2 E_{ps}, \qquad (9)$$

where the first and second terms are the power consumption related to the optical signal and phase shifters, respectively. Note that this power consumption was evaluated when the MRR crossbar array works with $TOPS$ calculated using (8). $N$ and $N^2$ in the equation correspond to the numbers of input/output ports and phase shifters in the PIC, respectively. In the first term, the power consumption related to sending and receiving optical signals per MVM $E_{signal}$ is assumed to 6.6 pJ/clock, which contains 2.1 pJ/clock for an analog-to-digital converter[4], 3.5 pJ/clock for a digital-to-analog converter[5], and 1 pJ/clock for other systems of optical interconnection. With the clock frequency, the power required for the optical signal increases. In the second term, $E_{ps}$ is the power consumption of a phase shifter for a $\pi$ phase shift. Since we consider the benchmark of the inference task, the injection currents of the phase shifters do not change frequently, and we can consider $E_{ps}$ under static condition. We consider a TO phase shifter and a III-V/Si hybrid metal-oxide-semiconductor (MOS) optical phase shifter[6]. In the case of the typical TO phase shifter, $E_{ps} = 20$ [mW]; in the case of the hybrid MOS optical phase shifter, $E_{ps} = 10$ [pW] owing to its operation principle[7]. Figure S16(b) shows the calculated relationship between $N$ and $E_{MVM}$ when $f_{CLK} = 3$ [GHz]. The red and blue lines correspond to the hybrid MOS optical and TO phase shifters, respectively. As shown in (6), the phase shift required for switching an MRR changes with $O(N^{-1})$. Thus, $E_{MVM}$ is proportional to $N$.

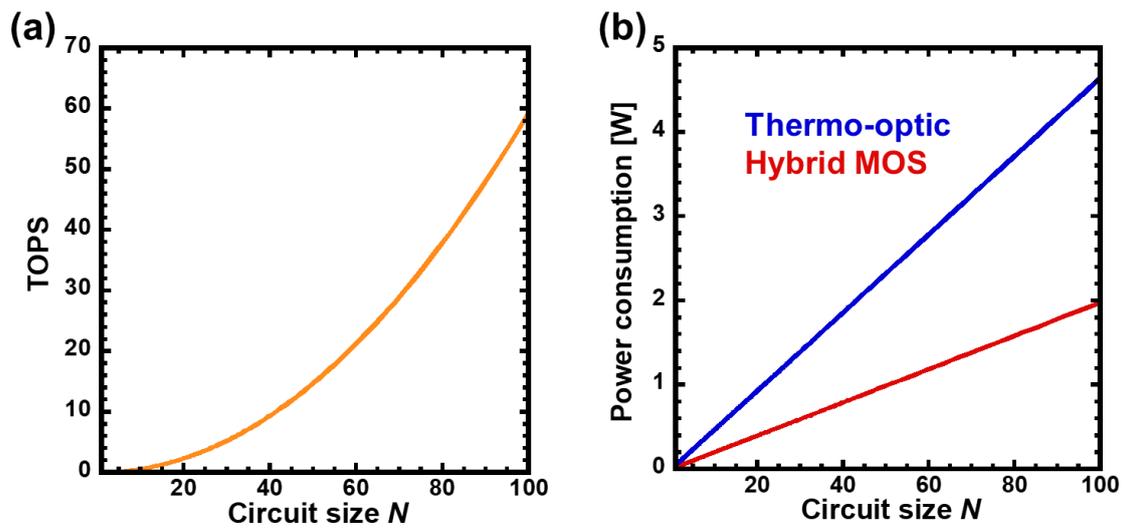

**Fig. S16.** Calculated (a) computation speed and (b) power consumption of MRR crossbar array as a function of its circuit size.